\documentclass[12pt, preprint]{aastex}
\usepackage{geometry} 

\slugcomment{accepted to AJ}

\shorttitle{Cyclic Variability of HDE 326823} 
\shortauthors{Richardson et al.} 

\begin{document}

\title{A Binary Orbit for the Massive, Evolved Star HDE 326823,\\ 
a WR+O System Progenitor}
\author{N. D. Richardson\altaffilmark{1,2}, 
D. R. Gies\altaffilmark{2}, and
S. J. Williams\altaffilmark{2}}

\altaffiltext{1}{Visiting astronomer, Cerro Tololo Inter-American Observatory, 
National Optical Astronomy Observatory, which are operated by the 
Association of Universities for Research in Astronomy, 
under contract with the National Science Foundation}
\altaffiltext{2}{Center for High Angular Resolution Astronomy, 
Department of Physics and Astronomy, 
Georgia State University, P. O. Box 4106, Atlanta, GA  30302-4106; 
richardson@chara.gsu.edu, gies@chara.gsu.edu, swilliams@chara.gsu.edu} 

\setcounter{footnote}{2}

\begin{abstract}
The hot star HDE~326823 is a candidate transition-phase object
that is evolving into a nitrogen-enriched Wolf-Rayet star.  
It is also a known low-amplitude, photometric variable with a 6.123~d period.  
We present new, high and moderate resolution spectroscopy 
of HDE~326823, and we show that the absorption lines show coherent 
Doppler shifts with this period while the emission lines display
little or no velocity variation.  We interpret the absorption line
shifts as the orbital motion of the apparently brighter star in 
a close, interacting binary.  We argue that this star is losing mass
to a mass gainer star hidden in a thick accretion torus and 
to a circumbinary disk that is the source of the emission lines. 
HDE~326823 probably belongs to a class of objects that produce 
short-period WR+O binaries. 
\end{abstract}

\keywords{stars: evolution
--- stars: individual (HDE 326823)
--- stars: winds, outflows
--- stars: Wolf-Rayet 
--- binaries: spectroscopic }

\section{Introduction}

Mass loss plays a key role in the evolution of the most massive stars, where, 
for example, a 60 $M_{\odot}$ star may lose 65\% of its mass before exploding 
as a supernova (Smith \& Owocki 2006). Very massive stars ($\gtrsim 25 M_{\odot}$; 
Meynet \& Maeder 2003) start their lives as O stars and will lose large amounts 
of mass as they pass through stages such as the red supergiant (RSG) and Wolf-Rayet (WR) 
phases during their post-main sequence evolution. 
The most massive stars may experience the rare, luminous blue variable (LBV) stage 
that is observed in objects such as P~Cyg and $\eta$~Car. During the LBV phase, 
a star is subject to possible great eruptions, variations on timescales from days 
to decades, and extremely high mass loss rates (Smith et al.\ 2011a). 

The loss of a star's envelope to become a WR star may occur entirely by winds 
(Massey, Conti, \& Niemela~1981) or it may be aided by Roche lobe overflow in a close binary 
to form WR+O systems like $\gamma^2$~Vel (North et al.\ 2007).  
The prototypical examples of the active mass-transfer objects are 
$\beta$~Lyr (Zhao et al.\ 2008) and the more massive RY~Sct (Grundstrom et al.\ 2007). 
In both of these cases, we observe the less massive donor star while 
the more massive gainer star is hidden within an optically thick, accretion torus. 
Such H-deficient, donor stars will likely explode in Type~Ib/c supernovae (Smith et al.\ 2011b),  
and understanding their pre-SN evolution is critical to the interpretation and modeling of these supernovae. 

HDE 326823 (Hen 3-1330; V1104 Sco; ALS 3918; ASAS J170654-4236.6) is 
a Galactic example of an unusual transition-phase object. 
Van Genderen (2001) places this object in the category of ex-/dormant LBVs, 
and the star may be entering the WN (nitrogen-enriched WR) stage of evolution 
from either the LBV or RSG stage (Lopes et al.\ 1992; Sterken et al.\ 1995a; 
Marcolino et al.\ 2007).  Sterken et al.\ (1995a) found some changes in 
Str\"{o}mgren photometry over long timescales from data taken three years apart, 
and they presented some examples of line profile variations in the optical spectrum. 
Analysis of the more recent photometric variations from the 
All Sky Automated Survey\footnote{http://www.astrouw.edu.pl/asas} (ASAS) 
by Pojma\'{n}ski \& Maciejewski (2004) indicates a short period of 6.123~d and 
an amplitude of 0.17~mag in $V$. 

The far-UV spectrum of HDE~326823 resembles that of an early B-supergiant
(Shore et al.\ 1990).  However, the optical spectrum of HDE 326823 was described as 
``the most bizarre in the OB Zoo'' by Walborn \& Fitzpatrick (2000), 
who discussed the weak H lines, the strong \ion{Fe}{2} emission, and 
several other subtle features.  They placed this star in the category of an 
``iron star'' from the presence of the \ion{Fe}{2} emission, a group 
that includes only four other high luminosity objects in their study. 
The unusual spectrum of HDE~326823 was further examined in two key studies. 
Borges Fernandes et al.\ (2001) presented a spectral atlas for the region 
between 3800 and 9200 \AA.  They found very strong helium emission lines 
while the hydrogen Balmer emission lines were relatively weak. 
The \ion{He}{1}, \ion{H}{1}, and \ion{Ca}{2} emission lines all exhibited 
double-peaked profiles.  Marcolino et al.\ (2007) performed detailed modeling 
of the spectrum from the ultraviolet through the optical with the non-LTE 
radiative transfer code, CMFGEN (Hillier \& Miller 1998). 
They found that the star has wind parameters that are similar to those of 
low luminosity LBVs.  Most importantly, they found that the mass fractions are 
$X_H \sim 3\%$ and $X_{He} \sim 96\%$, and their analysis supports the 
interpretation that the object is a pre-WN star.  Their CMFGEN model fits the 
spectral energy distribution and most emission lines, but does not account for 
the double-peaked profiles.

In this paper, we present in \S2 a re-analysis of the photometry from ASAS, 
and we show that the complex ASAS light curve is consistent with the 
earlier photometric variations observed by Sterken et al.\ (1995a). 
In \S3, we present new, moderate and high resolution, spectroscopic observations,
use radial velocity measurements of weak \ion{N}{2} absorption features to 
derive orbital elements based upon the photometric period, and show 
how the emission profiles vary with this period.  We develop in \S4 a binary scenario 
to explain the observational properties.  We summarize our results in \S5.

\section{Photometric Variations}

We re-analyzed the light curve by first collecting 
$V$-band measurements from ASAS (Pojma\'{n}ski 2002) by removing points deemed lower quality. 
We performed a time-series analysis on the remaining 479 $V$-band measurements spanning 3155~d using the 
Lomb-Scargle periodogram (Scargle 1982). In the period domain of 2--40~d, 
we find strong peaks around periods 3 and 6~d. However, the 3~d period is 
a sub-multiple of the main period of photometric variability of 6.1228$\pm$0.0053~d 
(Fig.~1, top), confirming the 6.123 d period listed in the ASAS Catalog of 
Variable Stars (Pojma\'{n}ski \& Maciejewski 2004).  We see two distinctive 
minima and maxima in the phased light curve (maxima near phase 0.35 and 0.9 and 
minima around phase 0.1 and 0.8, when phased to an epoch of HJD 2454622.4312, 
introduced in \S3.3).  Pojma\'{n}ski \& Maciejewski (2004) classify the 
light curve as a DCEP-FU type, i.e., that of a Cepheid pulsating in the fundamental mode. 
However, this star is so far removed from the Cepheid instability strip in the 
H-R diagram that we must seek other explanations (\S4).

We also examined the archival measurements of the Long-term Photometry of Variables 
(LTPV) program of Str\"{o}mgren $y$ magnitudes (Sterken et al.\ 1993; Manfroid et al.\ 1995; 
Sterken et al.\ 1995b; star ID P7058).  Sterken et al.\ (1995a) analyzed these data and found a 
systematic brightening between data collected from 1988 and 1991 and data collected in 1994. 
We show their $y$-band measurements in the second and third panels of Figure~1, 
where the 1994 data are plotted in the middle panel and the earlier measurements are 
shown in the bottom panel.  These light curves are qualitatively similar to the 
ASAS light curve, but appear shifted in phase.  The appropriate phase shift was 
found by cross-correlating in phase the observed set with a phase-binned,  
averaged light curve from ASAS.  The shifted ASAS light curve is shown in 
the middle and bottom panels of Figure~1 for phase shifts of $\Delta \phi = +0.08 \pm 0.02$
and $\Delta \phi = +0.28 \pm 0.02$, respectively.  These shifts indicate that
the actual period may be slightly smaller and/or was smaller in the past compared
to the period derived from ASAS, and we encourage continued photometric observations. 
Nevertheless, the reasonable agreement between the shifted ASAS light curve and 
the LTPV data sets indicates that the complex, cyclic, light variations have likely 
been present over the last two decades.

\section{Spectroscopic Variations}

\subsection{Observations}

We obtained 21, low dispersion, red spectra of HDE~326823 with the 
CTIO 1.5~m telescope and Cassegrain R-C spectrograph (SMARTS Consortium setup 47/Ib; 
described by Howell et al.\ 2006). These spectra have a resolution of 2.2 \AA\ FWHM 
($R \simeq 3000$).  The spectra were recorded on a Loral 1200$\times$800 CCD detector 
and were reduced using standard techniques.  We typically made two 120~s integrations 
per visit and performed the wavelength calibration using a Ne lamp, with resulting 
residuals typically of 0.02 \AA. 

We also obtained high resolution spectroscopy ($R\sim 40000$) with the CTIO 1.5~m 
telescope and the fiber-fed, echelle spectrograph\footnote{http://www.ctio.noao.edu/$\sim$atokovin/echelle/FECH-overview.html}.  
We typically made two 900~s integrations per visit. The resulting echelle spectra 
are of good quality in the red-green region, where we typically obtained a 
signal-to-noise ratio of 50--75 per pixel in the continuum.  
These spectra were reduced using standard echelle spectroscopy techniques with 
IRAF\footnote{IRAF is distributed by the National Optical Astronomy Observatory, 
which is operated by the Association of Universities for Research in Astronomy, Inc., 
under cooperative agreement with the National Science Foundation.}, 
but the spectra were flat-fielded after extraction with custom IDL software 
in order to remove residual jumps in the intensity across the center of the dual-readout chip. 
The wavelength solution was determined by means of a Th-Ar lamp spectrum that 
was made at the same sky location just prior to the science exposure. 
Typical residuals to the fit are on the order of 0.006 \AA. 
We obtained spectra over five nights during the commissioning period for the spectrograph 
(2008 Jun 5--9) and over 15 additional nights during 2010. 
The commissioning period spectra cover a range 4780--6450 \AA, while the more recent spectra 
were collected with the full chip being read out and cover a range of 4200--7350 \AA. 
Note that all of these data suffer from very low signal-to-noise at regions blueward of $\sim 5500$ \AA. 
A few of these spectra were omitted in the analysis, because of their low signal-to-noise ratio.

\subsection{Absorption Line Variability}

Very few absorption lines are present in the red part of the spectrum of 
HDE 326823 (Borges Fernandes et al.\ 2001), but there are a few \ion{N}{2} lines 
in the region of 5667--5710 \AA\ that are recorded in our echelle observations
(see Fig.~1b in the paper by Borges Fernandes et al.\ 2001). 
In Figure 2 we show a two panel plot of these features as a function of 
photometric phase, both as line plots (top panel) and as a gray scale 
that is interpolated in phase to fill in observational gaps (lower panel). 
There is a subtle, double-peaked, emission feature blueward of these \ion{N}{2} lines 
that is identified as \ion{Fe}{2} $\lambda\lambda 5657, 5659$ by Borges Fernandes et al.\ (2001). 
The \ion{N}{2} absorption lines show large radial velocity shifts as a function of photometric phase, 
while the \ion{Fe}{2} emission remains kinematically stationary.

These \ion{N}{2} features are typically observed in the spectra of B-type supergiants. 
We computed a reference spectrum by using the values of $T_{\rm eff} = 22,280$ K and $\log g = 2.9$ 
from the spectral modeling of Marcolino et al.\ (2007) and by interpolating within the 
BSTAR2006 grid of model spectra (Lanz \& Hubeny 2007). This reference spectrum is shown 
in the top panel Figure~2 with the continuum placed at phase 1.1. The \ion{N}{2} lines 
in the model match reasonably well with those in the observed spectrum in most cases, 
but there were at least four epochs where these lines weaken or disappear entirely. 

\subsection{A Single-Lined Spectroscopic Orbit}

We measured radial velocities for the \ion{N}{2} lines by means of cross-correlation 
with the reference spectrum computed above. Our results produced reliable velocities for all but 
four observations where the \ion{N}{2} lines were too weak to measure. 
The average statistical error of the cross-correlation velocities is 6.6 km~s$^{-1}$
according to the method of Zucker (2003), but the individual errors are much
larger in those cases where the \ion{N}{2} lines are weak. 
We used these measurements and the derived photometric period of 6.1228~d 
to make an error-weighted fit of the remaining orbital elements using 
the program of Morbey \& Brosterhus (1974).  The measurements are given in Table~1, 
and the derived, single-lined, orbital elements are shown in Table~2
(where $T$ is the epoch of periastron that we adopted in \S2 for the light curve). 
We plot in Figure~3 the radial velocities as a function of orbital phase, 
together with the derived orbital velocity curve.

Our derived orbit has two features that are striking. 
The first is that an eccentric orbit fits the data much better than a circular orbit. 
When we fit a circular orbit, the rms of the residuals from the fit increased by more than a factor of two. 
A significant eccentricity is unusual for a short period binary because the large tidal forces should 
act to circularize the orbit.  The second notable aspect of this orbit is that 
the mass function is large, with a value of $7.3 M_{\odot}$.  This suggests that the companion must be a massive 
star, yet there is no clear evidence of spectral features with Doppler shifts
corresponding to those of the companion.  We will discuss the nature of the 
companion star in \S4.

\subsection{Emission Line Variability}

\subsubsection{\ion{He}{1} Lines}

Some of the strongest emission features in the optical spectrum of HDE~326823 are the 
\ion{He}{1} $\lambda\lambda 5876, 6678, 7065$ lines (Borges Fernandes et al.\ 2001).
We show the \ion{He}{1} $\lambda 5876$ profiles in Figure~4 as a function of 
radial velocity and orbital phase.  The emission line appears relatively constant
and is always double-peaked, similar to those observed in the spectra of Be stars 
where the flux originates in a circumstellar disk (Porter \& Rivinius 2003). 
However, to emphasize the subtle variations related to phase, we subtracted 
the average profile from the observations to produce a phase-interpolated, 
grayscale depiction of the difference spectra that appears in the lower panel of Figure~4.
The difference spectra exhibit an absorption subfeature that moves in the same
way as found for the \ion{N}{2} absorption lines (the \ion{N}{2} radial velocity
curve is overplotted as a white line).   We interpret this as the photospheric 
\ion{He}{1} $\lambda 5876$ absorption line of the visible star. 
In \S4 we discuss several other subtle, phase-related variations, such as the well-defined, 
red edge of the emission that first appears near phase 0.7 and progresses linearly 
blueward until near phase 0.4.  Similar line and grayscale difference plots 
for \ion{He}{1} $\lambda\lambda 6678, 7065$ are shown in Figures 5 and 6, respectively, 
and these show variations similar to the case of \ion{He}{1} $\lambda 5876$. 

Remarkably, the bulk of the emission in \ion{He}{1} $\lambda 5876$ shows no evidence of 
the orbital Doppler shifts we found for the \ion{N}{2} absorption lines (\S3.3). 
We checked this absence of motion by measuring the radial velocity of the steep line wings 
of \ion{He}{1} $\lambda 5876$ using a line bisector method (Shafter et al.\ 1986). 
The bisector velocities are determined by cross-correlating the observed profile 
with a template formed of oppositely signed Gaussian functions with FWHM = 75 km~s$^{-1}$ 
at offset positions of $\pm 150$ km~s$^{-1}$.  The zero-crossing of the resulting
cross-correlation function yields the bisector velocities $V_r$ that 
are listed in Table~3 and plotted as plus signs in Figure~3.  
Typical uncertainties for these velocities are $\pm 3$ km~s$^{-1}$ for the echelle data and 
$\pm 24$ km~s$^{-1}$ for the lower resolution Cassegrain spectra. 
We see a small amplitude variation in the opposite sense to that of the 
\ion{N}{2} absorption lines that we attribute to the influence of the weak 
\ion{He}{1} absorption component that causes an apparent reduction in 
the emission wing flux at the Doppler shift extrema.   Thus, we conclude that 
the \ion{He}{1} $\lambda 5876$ feature is well explained by the superposition of 
a kinematically static emission line and an orbitally-modulated absorption 
component from the visible supergiant.

The other remarkable fact is the near constancy of the \ion{He}{1} $\lambda 5876$ 
emission strength indicating the emission source is never substantially
occulted by the stars over the course of the orbit.  We measured the net equivalent 
width of the feature by a numerical integration from $-500$ to $+1200$ km~s$^{-1}$
across the profile.  The larger, positive boundary necessarily 
includes the \ion{Na}{1} $\lambda 5890$ line because the 
resolution of the R-C spectrograph blends the red wing of \ion{He}{1} $\lambda 5876$ 
with the blue emission wing of \ion{Na}{1} $\lambda 5890$.  The entire Na~D component 
remains approximately constant and makes a negligible contribution to the variability 
of \ion{He}{1} $\lambda 5876$ (see Fig.~4).  The formal errors associated with these 
equivalent width measurements are $\pm 2\%$ for all of these data, primarily set by 
lower signal-to-noise in the echelle data and by uncertain continuum placement 
in the Cassegrain spectra (due to the large number of unresolved emission lines). 
These measurements are listed in Table~3 and are plotted in Figure~7. 
There is a systematic 1.1 \AA\ difference in the average equivalent width for 
the two data sets that is due to differences in continuum placement.
If the absolute emission flux remains constant throughout the orbit, then 
we would expect to observe an apparent change that varies as the inverse of
the continuum flux.  We overplot in Figure~7 scaled versions of the inverse of the flux 
from the smoothed ASAS $V$-band light curve, and we see that the equivalent width variations 
are similar to those expected for a constant emission line flux relative to a changing continuum.
This indicates that the emission source remains clearly visible throughout the orbit. 

\subsubsection{H$\alpha$}

HDE~326823 is extremely hydrogen deficient with a H mass fraction of 3\% (Marcolino et al.~2007), 
and the only hydrogen lines observed in the optical spectrum are H$\alpha$ and H$\beta$
(Borges Fernandes et al.\ 2001).  We show in Figure~8 the phase-related behavior 
of H$\alpha$ observed in echelle spectroscopy from 2010.  The feature appears more or less constant
throughout the orbit, a property confirmed in the set of Cassegrain spectra. 

\subsubsection{Metal Lines}

There are several \ion{Fe}{2} emission lines in our spectra of HDE 326823 that are weak, 
double-peaked features.  In general, the S/N ratio of our data is too low to investigate 
the orbital-related variations in these lines, but we show one example in Figure~9. 
This emission feature is identified by Borges Fernandes et al.\ (2001) as a blend of 
\ion{Fe}{2} $\lambda 6345$, \ion{Si}{2} $\lambda 6347$, and \ion{Ni}{2} $\lambda 6347$. 
We see some evidence of changes in the blue and red emission peak strength, and  
the difference profiles shown in the lower panel suggest a progressive flux shift 
from the red to blue emission extremes between orbital phases 0.3 and 0.9. 
Our interpretation will be discussed in \S 4.

\section{Discussion}

The facts that have emerged from our spectroscopic analysis are striking and challenging. 
HDE~326823 is a short period binary with a massive, yet unseen companion. 
The low H abundance of the visible star (Marcolino et al.\ 2007) indicates that 
it has lost a significant fraction of its outer envelope, and the presence of 
strong emission lines indicates the presence of circumstellar gas from 
ongoing mass loss.  All these properties suggest that HDE~326823 is an 
interacting binary which is experiencing active mass transfer.  
Here we propose that HDE~326823 is related to the W~Serpentis class of 
massive binaries (Tarasov 2000), systems in which the mass donor appears 
as the visible, lower mass component and the more massive, mass gainer is hidden 
behind a thick accretion torus (Nazarenko \& Glazunova 2006). 

The mass function $f(M)$ (Table 2) indicates that the total mass
of the system is large, 
$$M_1 + M_2 = 7.3 M_{\odot} \left(1 + {1\over q}\right)^3 \sin ^{-3} i$$
where the mass ratio is $q=M_2 / M_1$ and the subscripts 1 and 2 denote the visible 
and unseen companion star, respectively.  Both the terms involving
$q$ and $i$ on the right hand side of the expression are greater than 
unity, so the total mass is potentially very large.  On the other hand, 
Marcolino et al.\ (2007) found that the total luminosity of HDE~326823 
corresponds to that of a single star with an initial mass of $25 M_\odot$, 
and this suggests that the mass ratio term should be relatively small 
or, equivalently, $q > 1$.  This agrees with our expectation that the 
visible mass donor has lost most of its mass to the gainer and that the mass 
ratio has reversed.  We can make an initial estimate of the mass ratio
by comparing the projected rotational velocity $V \sin i$ to the 
orbital semiamplitude $K_1$ (Gies \& Bolton 1986), 
$${{V \sin i}\over {K_1}} = \rho ~\Omega \left(1 + {1\over q}\right) \Phi (q).$$
This expression relates the size of the visible star 
to the Roche radius $\Phi (q)$ (Eggleton 1983) 
through a fill-out factor $\rho$ ($=1$ for a Roche filling star), and the angular
rotational rate is expressed relative to the synchronous rate through factor $\Omega$. 
Based upon the absorption line profiles published by Borges Fernandes et al.\ 
(2001; see their Fig.~5), we estimate that $V \sin i = 83 \pm 15$ km~s$^{-1}$ (or smaller if the line broadening is significantly influenced by macroturbulence) from a visual estimate of the FWHM of the profile. 
We assume that the visible star fills its Roche lobe at periastron 
(based upon evidence of mass transfer and mass loss), so that the average fill-out
factor is $\rho = (1 - e) = 0.81 \pm 0.06$.  Furthermore, we imagine that tidal forces 
have acted to force the star into the synchronous rotation rate that occurs
at periastron (when tides peak), and therefore, $\Omega = (1 - e)^{-2} = 1.51 \pm 0.22$.
Then the expression above can be solved for the mass ratio, $q= 5.3^{+3.9}_{-1.8}$, 
which is again consistent with the mass ratio reversal expected for the 
advanced stage of evolution of this binary. 

We present a graphic representation of the system geometry in Figure~10 that 
shows the binary as viewed from above the orbital plane.  
For the purposes of this diagram, we adopt $q= 5.3$ and $i=45^\circ$, 
which leads to masses of $M_1=5.5 M_\odot$ and $M_2=29.1 M_\odot$. 
The central binary is drawn to a scale for the time of minimum separation at periastron.
The visible star (left) is assumed to fill its Roche surface at periastron,
and the companion star (right) is shown with a radius of $10 R_\odot$, 
appropriate for a main sequence star with the adopted mass (Martins et al.\ 2005).  
Arrows centered in each star represent their orbital velocity
about the center of mass (indicated by a tick mark within the figure of the companion), 
and the numbers in each quadrant surrounding the binary give the orbital 
phase corresponding to the direction to the observer.

We suggest that many of the properties of HDE~326823 can be understood 
in the context of the mass transfer and mass loss processes observed 
in other W~Ser systems such as $\beta$~Lyr (Zhao et al.\ 2008) and
RY~Sct (Grundstrom et al.\ 2007).  The visible star in such systems
is losing mass by Roche lobe overflow (RLOF) to the companion.  
Mass transfer is accompanied by angular momentum transfer that 
can spin-up the companion to the critical speed where it can no 
longer easily accrete additional gas.  The gas then accumulates in 
a thick torus that surrounds and obscures the companion star. 
The RLOF is indicated in Figure~10 by thick and thin 
stream lines between the stars, while the torus is shown as a shaded region 
surrounding the companion (out to an assumed radius of $80\%$ of the 
companion's Roche lobe).  The large optical depth of the torus gas 
is the probable explanation for the absence of the companion's spectral 
features in the observed spectrum.   

Hydrodynamical models by Nazarenko \& Glazunova (2006) suggest that mass loss 
from the binary can occur through the loss of torus gas through the L3 region 
and through the lower gravitational potential region around L2 on the 
far side of the mass donor.  We show in Figure~10 example trajectories
of the gas leaving the donor through the L2 region (left of mass donor). 
These were calculated in the restricted three-body approximation ignoring 
the small eccentricity of the orbit.  This mass stream feeds a circumbinary disk
(outer shaded region in Fig.~10) with an inner radius of $2.83 a$ (where 
$a$ is the semimajor axis) that corresponds to the innermost stable orbit
according to the calculations of Pichardo et al.\ (2008).

We suspect that the emission lines in the spectrum of HDE~326823 form 
in the circumbinary disk of the system.  This location is consistent with
the lack of observed binary orbital motion, the large volume of formation, 
cooler gas conditions, and Keplerian motion that are required to explain 
the double-peaked emission profiles of H$\alpha$ and the \ion{He}{1} and 
\ion{Fe}{2} lines.   The projected Keplerian rotational velocity of 
circumbinary disk gas as a function of distance $r$ from the binary is
given by 
$$V_K \sin i = {{2 \pi a \sin i}\over P}~ \left({r\over a}\right)^{-1/2}   
 = 226~ {\rm km~s}^{-1} \left(1 + {1\over q}\right) \left({r\over a}\right)^{-1/2}.$$
Thus, the highest emission speed we would expect in the model 
for circumbinary gas would occur for the inner disk boundary 
at $r/a = 2.83$, or $V_K \sin i = 160$ km~s$^{-1}$ (for $q = 5.3$). 
This is consistent with the \ion{He}{1} $\lambda 5876$ profiles 
that have an average half-width at half-maximum of 146 km~s$^{-1}$ (see Fig.~4).  

A circumbinary disk being fed through the L2 point would explain 
two observed features in the \ion{He}{1} profiles.  
We see an excess of blue-shifted emission near phase 0.2 (Fig.~4). 
From our depiction of the binary (Fig.~10), this phase corresponds 
to times where the mass stream that is ejected from L2 appears 
in the foreground with negative radial velocity.  As the gas stream
moves away from L2 and lags behind the binary motion, the velocities 
within the outward spiral will become closer to the Keplerian speeds 
given above.  Thus, the fastest motions we observe at any given orbital 
phase will correspond to the projected gas speed in the part of spiral with 
the largest projected separation from the central binary.  Since this 
separation will increase and the associated Keplerian velocity decrease
with advancing phase, we might expect to observe a gradual decline
in the wing velocity that is similar to what we observe after phase 
0.0 in the blue wing and especially after phase 0.5 in the red wing of 
\ion{He}{1} $\lambda 5876$ (see Fig.~4). 
 

The binary scenario may also explain the lack of H$\alpha$ variability
(\S 3.4.2) and the orbital phase-dependent behavior of the metal lines
(\S 3.4.3).  We suggest that all these emission lines form in the
circumbinary disk, since they are generally constant in radial velocity.
Unlike the \ion{He}{1} emission lines, there is no apparent
photospheric component crossing H$\alpha$ (Fig.\ 8), and this may
be due to the low H abundance in the atmosphere of the mass donor.
We suspect that \ion{Fe}{2} formation is favored in the somewhat
cooler and dense gas conditions of the circumbinary disk, and the donor
star's atmosphere may be too hot to produce any photospheric \ion{Fe}{2}
absorption lines.  Indeed, Marcolino et al.\ (2007) found that
models for the star alone were characterized by higher ionization
states in the atmosphere (\ion{Fe}{3} and \ion{Fe}{4}), so
the presence of a somewhat cooler circumbinary disk offers
a reasonable explanation for the presence of the \ion{Fe}{2} emission lines.
The moving emission peak observed in \ion{Fe}{2} (Fig.\ 9)
may result from a gas density enhancement that occurs where the spiral
stream from L2 meets the inner boundary of the circumbinary disk (lower,
left part of the boundary in Fig.\ 10).  If the disk thickness and
inclination act to restrict our view to the far side of the inner cavity,
then we would expect to observe this over-dense part of the boundary with
maximum redshift near phase 0.4 and maximum blueshift near phase 0.9,
more or less in agreement with the emission shift observed in the
\ion{Fe}{2} difference profiles (Fig.\ 9).

We think that the binary model can also explain qualitatively the main 
features of the $V$-band light curve (Fig.~1).  Since the visible star 
fills or nearly fills its Roche surface, it will be tidally distorted 
and will display a characteristic ellipsoidal flux variation 
as we view the wide and narrow dimensions of the star with 
changing orbital phase.  According to the binary geometry shown in Figure~10, 
we expect to observe maxima near phases 0.97 and 0.43 and 
minima near phases 0.15 and 0.76, and these phases correspond
to the observed extrema in the light curve.  Furthermore, if the accretion 
torus is a significant contributor to the continuum flux, then changes
in the torus size with the elliptical orbit may affect the light curve. 
For example, increased mass transfer at periastron followed by increasing
binary separation to apastron may allow the optically thick torus to 
increase in size (to a radius shown by the dotted line in Fig.~10),  
and consequently the torus flux contribution would increase over the 
interval from phase 0.0 to 0.5.  This may explain why the maximum at 
phase 0.43 is brighter than the maximum at phase 0.97.  A detailed 
model of these flux variations should provide an estimate of the 
binary inclination, since the amplitude of the ellipsoidal variation 
varies as $\sin i$ while the amplitude of the torus size variation 
varies as $\cos i$ (in the thin disk limit).

We explored three other explanations of the periodic variations, 
but none of these were satisfactory in the end.  
First, suppose that HDE~326823 is a triple system consisting of 
a single LBV star with a distant 6~d binary.  This picture agrees
with the radial velocity constancy of the emission lines, 
but it is inconsistent with the apparent changes in the 
emission line wing morphology with orbital phase (Figs.~4, 5, and 6). 
Second, perhaps the emission lines originate in the accretion
torus surrounding the mass gainer star and the mass ratio is so 
extreme that the gainer's orbital motion is negligible. 
However, in this case the Keplerian gas motion in the torus would be 
very large because the gas is so close to the massive gainer 
star, and we would expect the emission line widths to exceed 
the semiamplitude of the binary companion.  This is opposite 
to our results that show that the binary semiamplitude $K_1$ ($230$ km s$^{-1}$)
is much larger than the characteristic half-width at 
half-maximum of the emission lines ($146$ km s$^{-1}$ for \ion{He}{1} 5876, with similar values for other species). 
Lastly, consider the possibility that HDE~326823 is a single star with 
a magnetically channeled wind that is observed at different 
orientations as the star rotates (Townsend \& Owocki 2005; Ud-Doula et al.\ 2009).  
Suppose that the magnetic and spin axes are slightly misaligned 
(the oblique rotator model; Shore \& Brown 1990) and that the outflow is observed 
in spectral lines as the magnetic pole transits the visible hemisphere. 
However, in the spectral lines of magnetically active massive stars, 
the wind material appears in absorption when its projected velocity 
falls within $\pm V \sin i$ and in emission at larger velocities. 
This is not the case in the spectrum of HDE~326823 where the \ion{N}{2} 
absorption features are Doppler shifted far beyond their characteristic line width. 

\section{Summary}

Our spectroscopic results demonstrate that HDE~326823 is a close, interacting 
binary with a 6.123~d orbital period that was originally found in ASAS light curve. 
We propose that the visible star (which has a spectrum similar to that of a B-supergiant) is the mass donor, 
and it is transferring mass to a more massive gainer star that is enshrouded 
in a thick accretion torus.  In addition, mass loss is occurring through both 
the L2 and L3 Lagrangian points into a large, circumbinary disk that is the 
source of the stationary emission lines.  The complex light curve probably results 
from a combination of the tidal distortion of the mass donor and variations
in the size of the accretion torus related to the elliptical orbit. 

HDE~326823 bears many similarities to other W~Serpentis interacting binaries
that are experiencing a phase of dramatic mass transfer.  However, its  
orbital period is shorter than that of most W~Serpentis binaries (with the 
possible exception of BD$+36^\circ 4063$ with $P = 4.8$~d; Williams et al.\ 2009),
and because mass transfer leads to an increasing period after mass ratio reversal, 
HDE~326823 may have emerged from its closest, most intense interaction. 
Further analysis of a long term light curve could yield a value for a 
changing period and yield an estimate of the mass transfer rate.
The spectral properties of HDE~326823 are perhaps most closely matched in the 
massive LBV MWC~314 that is also a binary with a 60.85~d period (Muratorio et al.~2008;
Lobel et al.\ 2012).  Lobel et al.\ (2012) find radial velocity variations in 
one set of absorption lines, while the double-peaked emission lines are constant 
in velocity, again suggesting a RLOF binary surrounded by a circumbinary disk.  

There are other known examples of luminous stars that are LBV candidates
and that display prominent \ion{Fe}{2} emission lines in their spectra 
(Walborn \& Fitzpatrick 2000; Massey et al.\ 2007), 
and some of these may in fact be binaries with 
circumbinary disks like HDE~326823.  However, we caution that 
the appearance of \ion{Fe}{2} emission only signifies the presence of 
relatively cool, dense, circumstellar gas, and such an environment might 
be attained in a dense stellar wind rather than in a circumbinary disk.  Verification of the similar 
binary nature of such targets will require high resolution, multiple-epoch spectroscopy 
to search for evidence of orbital Doppler shifts and to check for the 
presence of double-peaked and stationary emission lines.  

These new spectroscopic observations of HDE~326823 support the idea that 
some WR stars can be formed in a binary system through envelope loss by mass transfer. 
In this picture, the initially more massive star will expand to fill its Roche Lobe 
and begin mass transfer onto the smaller star.  
The mass donor will lose its hydrogen envelope and develop a WR spectrum. 
The mass gainer will emerge as an O-type star, with a large mass and high effective temperature. 
HDE~326823 may evolve into a system like $\gamma^2$ Vel (North et al.~2008), 
where the mass donor is now observed as a hydrogen stripped WR star, that is orbiting a more massive O star. 
Assuming that this binary history is correct, it is unlikely that the donor star 
in HDE~326823 experienced a RSG or LBV episode in the past since both kinds of objects 
have radii much larger than the current binary separation.  
HDE 326823 is likely experiencing a short, late stage of binary evolution, and 
it represents an important opportunity to explore the mass loss processes that occur at this juncture.

\acknowledgments 

We are grateful to Debra Fisher (Yale University) for collecting the 
first echelle spectra during the commissioning of the spectrograph. 
These spectra were made with the CTIO 1.5 m telescope, 
operated by the SMARTS Consortium.  We are grateful to 
Fred Walter (Stony Brook University) for his scheduling of this program, 
to the CTIO SMARTS staff for queue observing support, and 
to Todd Henry (Georgia State University) for assistance in scheduling 
the initial observations with the R-C spectrograph. 
Additionally, we thank Richard Townsend (University of Wisconsin) 
and Alex Lobel (Royal Observatory of Belgium) 
for suggestions that helped shape our conclusions. 
This research has made use of the SIMBAD database, operated at CDS, Strasbourg, France. 
This work was supported by the National Science Foundation 
under grants AST-0606861 and AST-1009080. 
Institutional support has been provided from the GSU College of Arts and Sciences 
and from the Research Program Enhancement fund of the 
Board of Regents of the University System of Georgia, administered through 
the GSU Office of the Vice President for Research. 
We gratefully acknowledge all this support.

{\it Facilities:} \facility{Cerro Tololo Inter-American Observatory's 1.5 meter Telescope} 



\begin{figure} 
\begin{center} 
\includegraphics[angle=90, width=15cm]{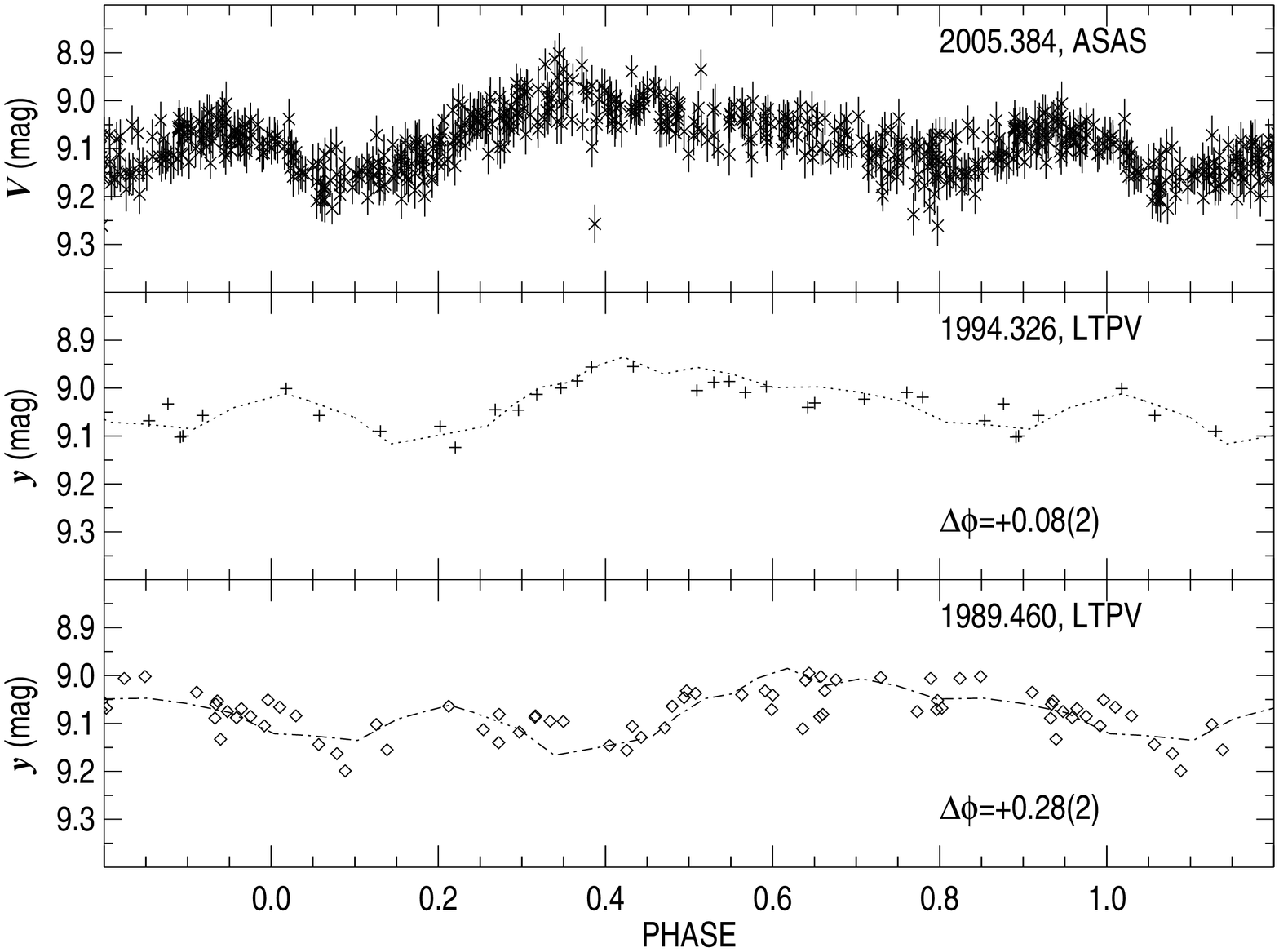}
\end{center} 
\caption{The ASAS $V$-band photometry is shown in the top panel according to 
the phasing described in the text. The middle and bottom panels illustrate the 
$y$-band measurements of Sterken et al.\ (1995a). Plus signs show data taken in 1994, 
while small diamonds represent data from 1988 to 1991.  The mean date of each set
is given in upper right part of each panel.  The phase-averaged (and shifted) 
ASAS light curve is overplotted for comparison (see text). The phase shift applied 
to the ASAS light curve is listed in the lower right part of the middle and lower panels
(with error estimates in the last digit quoted given in parentheses).} 
\label{fig1} 
\end{figure} 

\begin{figure} 
\begin{center} 
\includegraphics[angle=0, height=15cm]{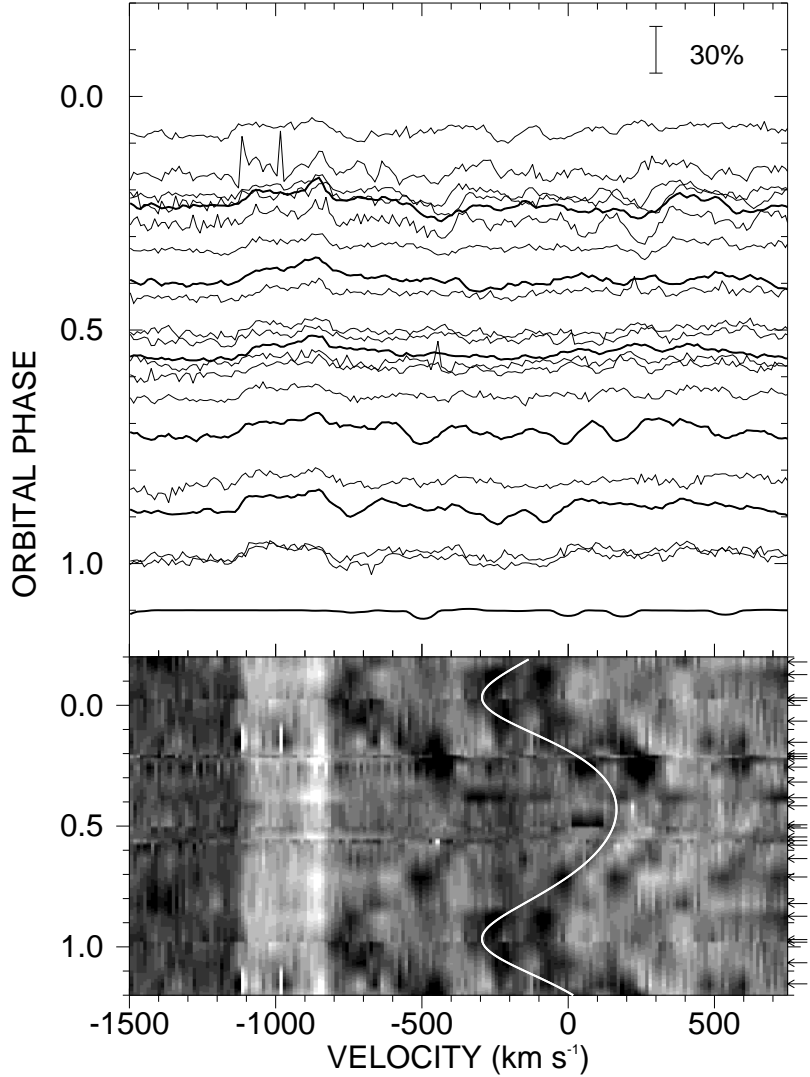}
\end{center} 
\caption{Spectra in the range 5648 -- 5690 \AA\ plotted as a function 
of heliocentric radial velocity (for $\lambda_0 = 5676.02$~\AA ) and
orbital phase.  This region includes \ion{Fe}{2} $\lambda 5673$ emission 
and \ion{N}{2} $\lambda\lambda 5666, 5676, 5679, 5686$ absorption lines.  
The top panel shows individual spectrum plots with the continuum set at the phase
of observation. The continuum scale is indicated with the vertical bar at the top.  Also depicted is a model 
reference spectrum (thick line; set with the continuum placed at phase 1.1). 
Spectra collected in 2008 are shown by thicker lines and those from 2010 by the thinner lines. 
The bottom panel shows a phase-interpolated, grayscale representation of the spectra. 
Arrows indicate phases of observations, and the white line shows the derived 
radial velocity curve in the reference frame for \ion{N}{2} $\lambda 5676$.} 
\label{fig2} 
\end{figure} 

\begin{figure} 
\begin{center} 
\includegraphics[angle=90, width=15cm]{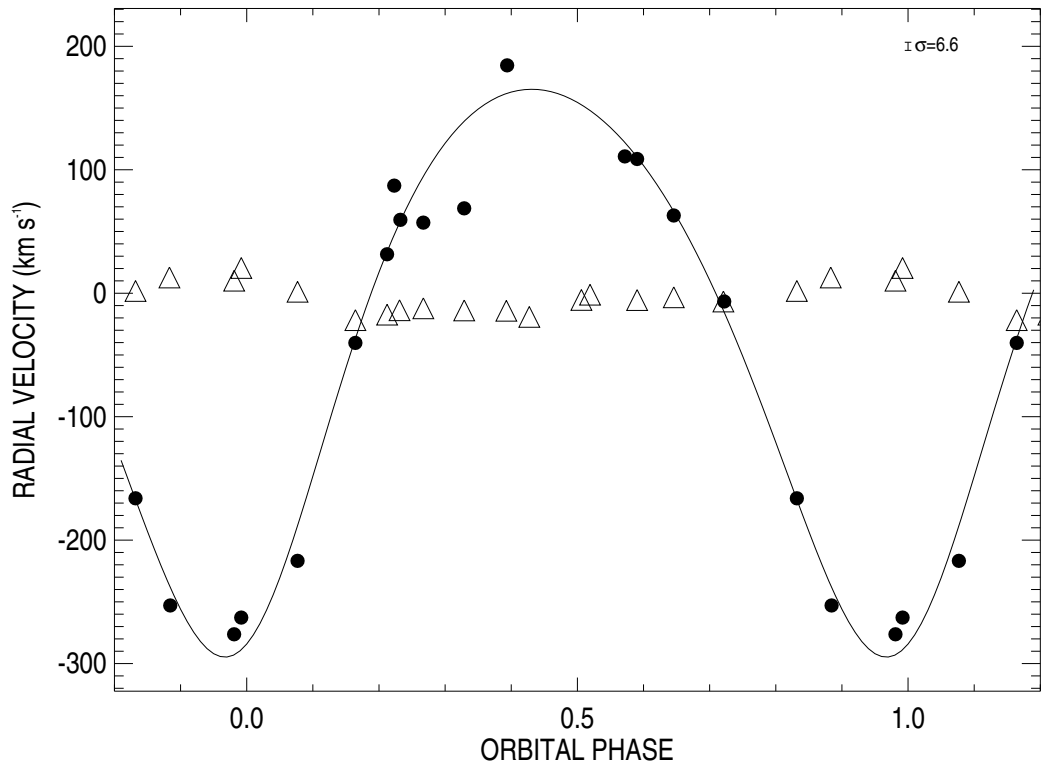}
\end{center} 
\caption{The orbital radial velocity curve (solid line) based upon the observed
\ion{N}{2} velocities (filled circles).  Overplotted as triangles are the radial velocities of the 
\ion{He}{1} $\lambda 5876$ emission line bisector. A typical statistical uncertainty for the \ion{N}{2} velocities (6.6 km s$^{-1}$) is shown in top right portion of the plot.} 
\label{fig3} 
\end{figure} 

\begin{figure} 
\begin{center} 
\includegraphics[angle=0, height=15cm]{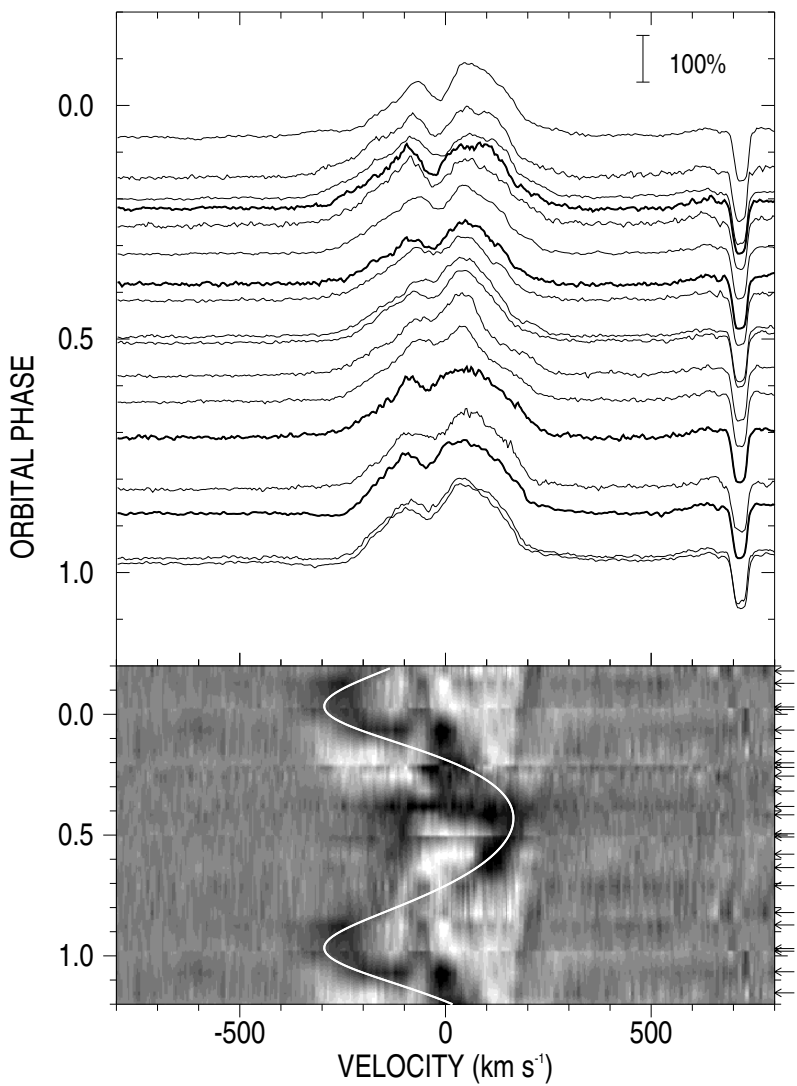}
\end{center} 
\caption{\ion{He}{1} $\lambda 5876$ echelle spectra shown as a function of heliocentric 
radial velocity and orbital phase in a similar manner as Fig.~2. The spectra from 2008 are shown 
as thicker line plots to show the long-lived phase dependent behavior. The bottom panel shows a grayscale depiction of the 
difference spectra, formed by subtracting the average profile from each individual spectrum. 
Arrows indicate times of observations, and the thick white line shows the derived 
orbital radial velocity curve from the \ion{N}{2} absorption lines. The variable 
absorption component appears to follow this velocity curve.} 
\label{fig4} 
\end{figure}

\begin{figure} 
\begin{center} 
\includegraphics[angle=0, height=15cm]{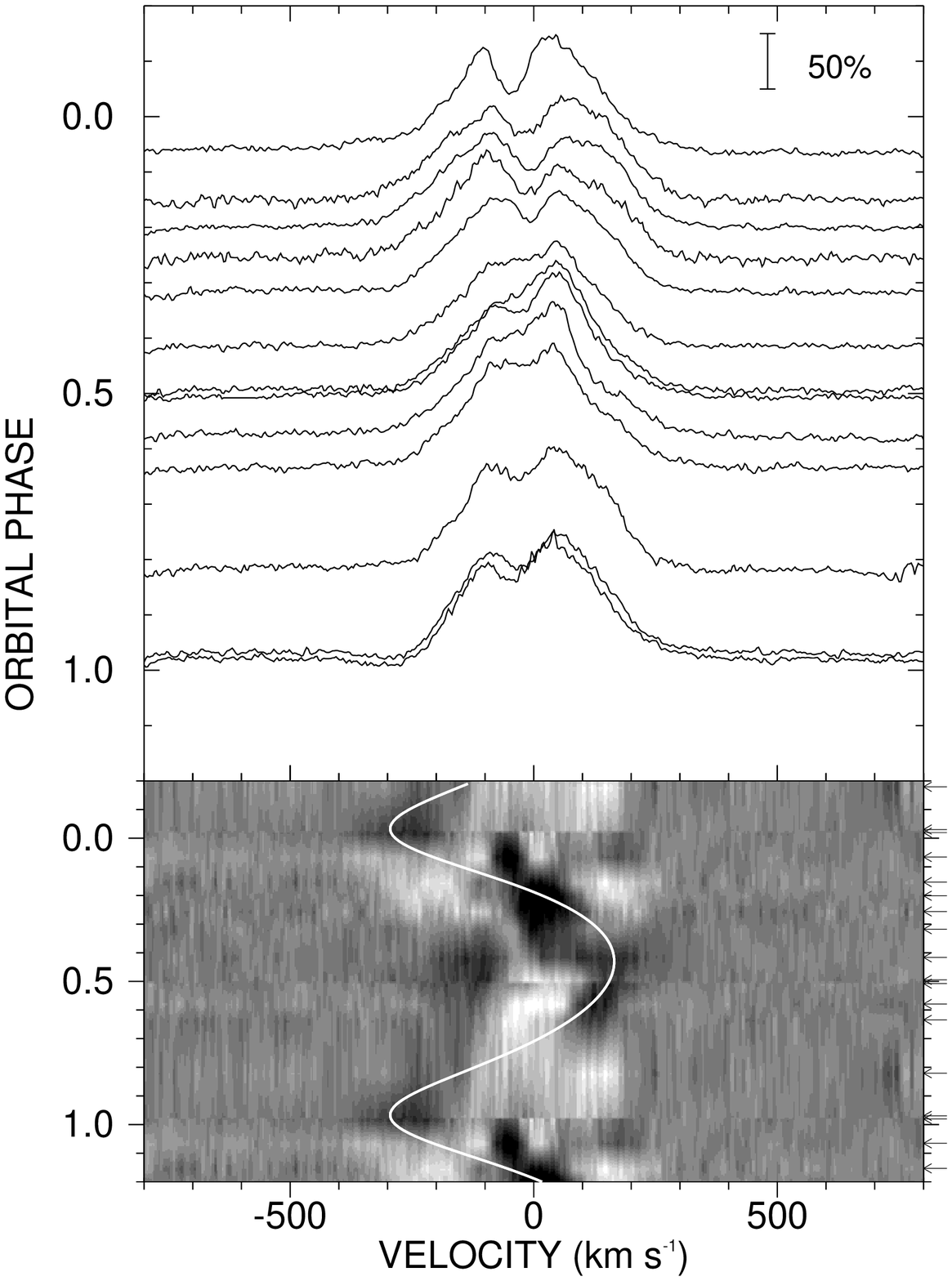}
\end{center} 
\caption{\ion{He}{1} $\lambda 6678$ echelle spectra (top) and difference spectra (bottom) 
presented in the same format as Fig.~4.} 
\label{fig5} 
\end{figure}

\begin{figure} 
\begin{center} 
\includegraphics[angle=0, height=15cm]{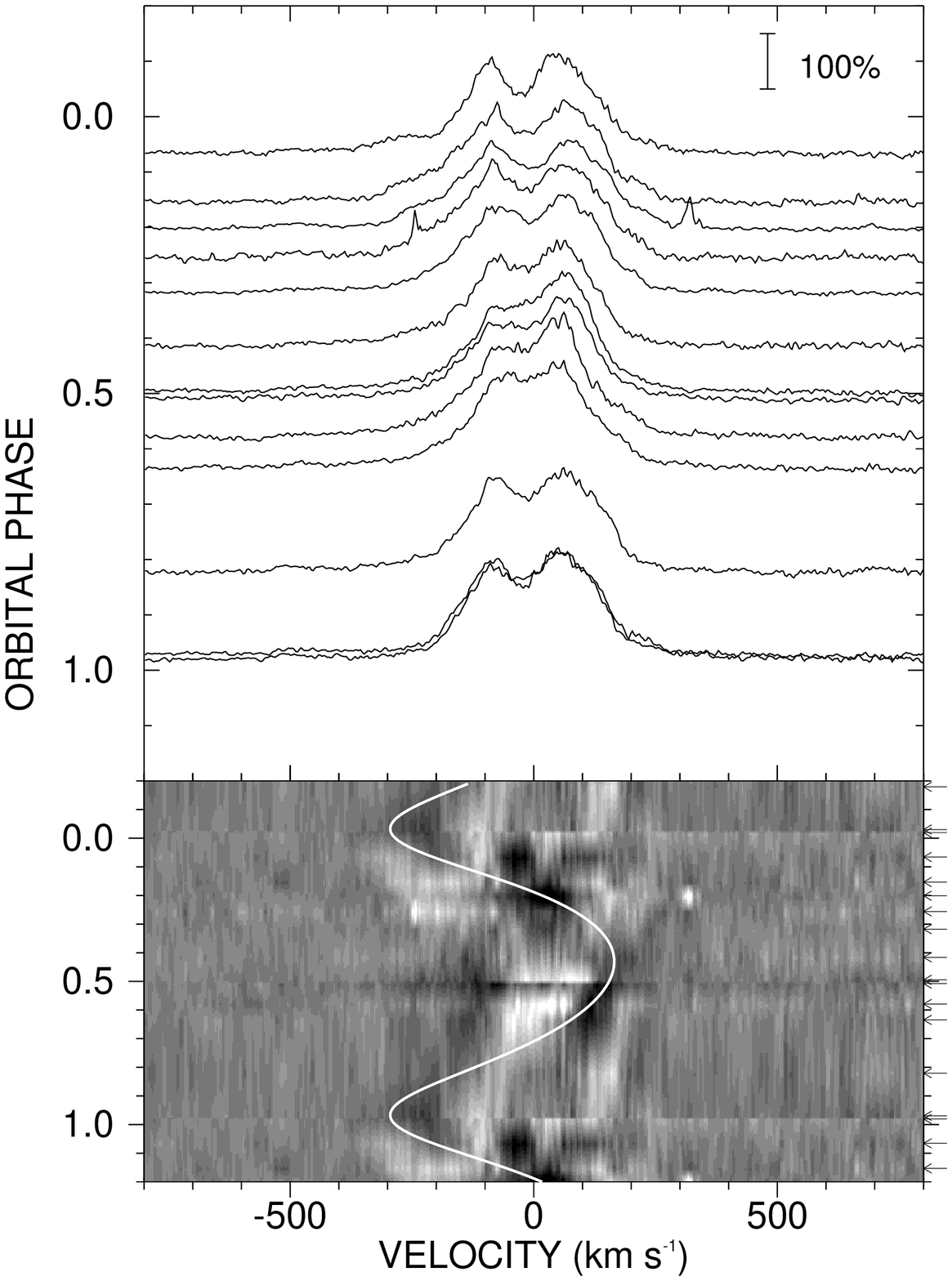}
\end{center} 
\caption{\ion{He}{1} $\lambda 7065$ echelle spectra (top) and difference spectra (bottom)
presented in the same format as Fig.~4.} 
\label{fig6} 
\end{figure}

\begin{figure} 
\begin{center} 
\includegraphics[angle=90, width=15cm]{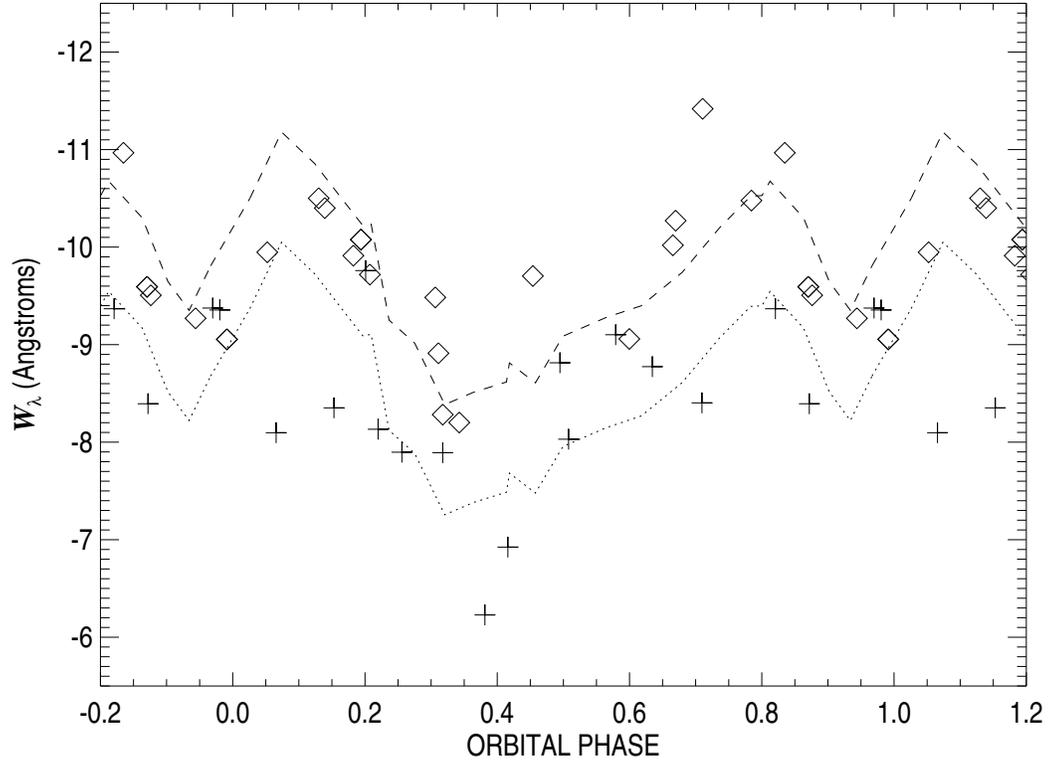}
\end{center} 
\caption{\ion{He}{1} $\lambda 5876$ equivalent width as a function of orbital phase. 
Measurements from the echelle spectroscopy are shown with $+$ symbols, and 
the Cassegrain spectra measurements are shown with $\diamond$ symbols. 
The dashed and dotted lines are predictions of the variations 
caused by the changing continuum flux, and they represent the 
inverse of the flux from the phase-averaged ASAS light curve 
that is scaled to the average equivalent width of each data set, with the dashed line corresponding to the Cassegrain measurements and the dotted line corresponding to the echelle measurements.} 
\label{fig7} 
\end{figure} 

\begin{figure} 
\begin{center} 
\includegraphics[angle=0, height=15cm]{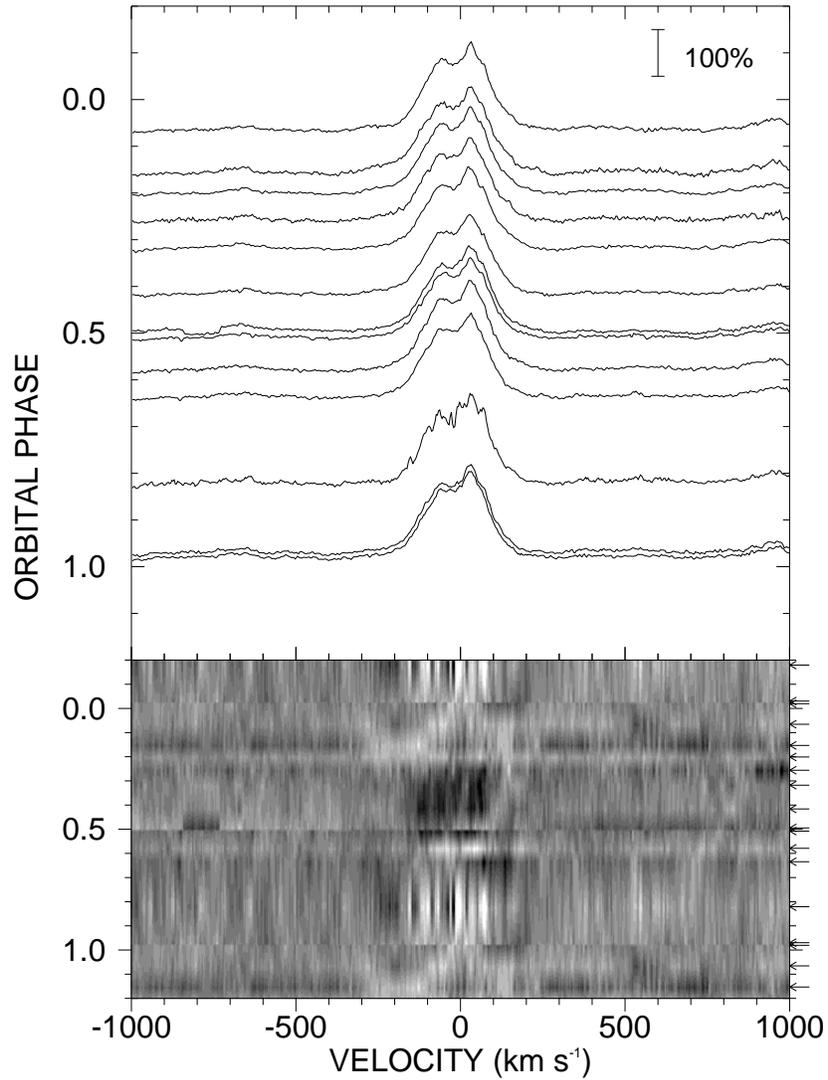}
\end{center} 
\caption{Echelle spectra of H$\alpha$ as a function of 
orbital phase and heliocentric radial velocity. 
The line plots (top panel) and grayscale representation of the difference spectra
(bottom panel) are in the same format as Fig.~4. No radial velocity curve is shown, as no photospheric component is observed.} 
\label{fig8} 
\end{figure}

\begin{figure} 
\begin{center} 
\includegraphics[angle=0, height=15cm]{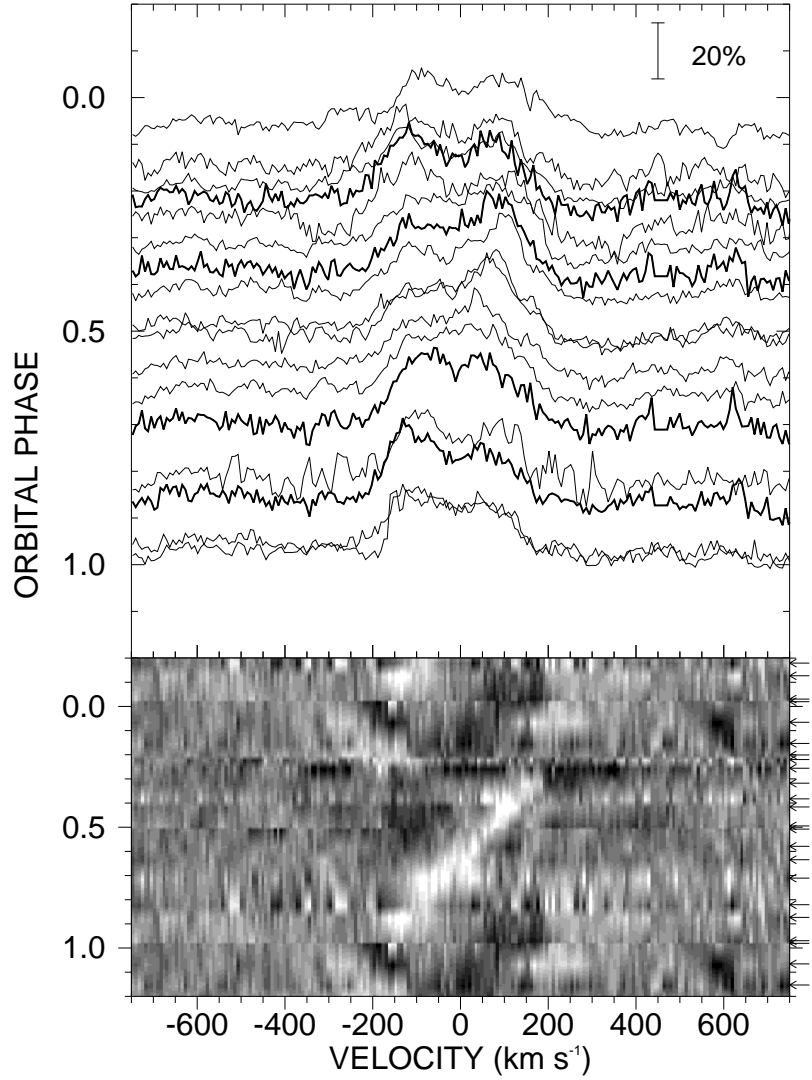}
\end{center} 
\caption{Echelle spectra of the \ion{Fe}{2} $\lambda 6345$, \ion{Si}{2} $\lambda 6347$, and 
\ion{Ni}{2} $\lambda 6347$ blend as a function of orbital phase and heliocentric 
radial velocity (in the reference frame of \ion{Si}{2} $\lambda 6347.11$). 
The line plots (top panel) and grayscale representation of the difference spectra
(bottom panel) are in the same format as Fig.~4. No radial velocity curve is shown, as no photospheric component is observed.} 
\label{fig9} 
\end{figure}

\begin{figure} 
\begin{center} 
\includegraphics[angle=90, width=15cm]{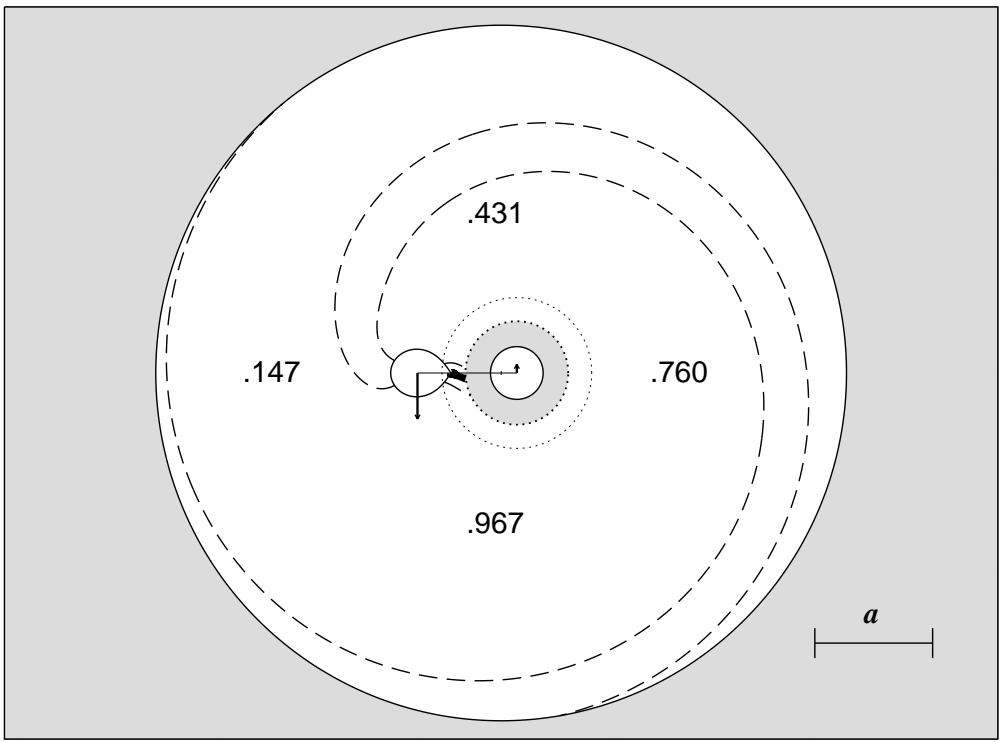}
\end{center} 
\caption{Graphic depiction of the binary system and circumbinary disk, as seen from above the orbital plane. See text for details.} 
\label{fig10} 
\end{figure} 


\begin{deluxetable}{cccc}
\tablecaption{\ion{N}{2} Radial Velocities
}
\tablewidth{0pt}
\centering
\tablehead{
  \colhead{Date}         &
  \colhead{Orbital}  &
  \colhead{$V_r$} & 
  \colhead{$\sigma(V_r)$} \\  
  \colhead{(HJD$-$2,450,000)}         &
  \colhead{Phase}  &  
  \colhead{(km s$^{-1}$)} &  
  \colhead{(km s$^{-1}$)}}  
\startdata
4623.8538	&	0.2323	&	$+$59.5\phn	 & \phn 5.7 \\
4624.8424	&	0.3938	&	$+$184.6\phn\phn & \phn 5.9 \\
4626.8519	&	0.7220	&	$-$6.7	         & \phn 4.0 \\
4627.8460	&	0.8844	&	$-$252.9\phn\phn & \phn 4.2 \\
5327.8527	&	0.2122	&	$+$31.6\phn	 & \phn 3.6 \\
5339.8048	&	0.1643	&	$-$40.2\phn	 & \phn 8.5 \\
5348.8753	&	0.6457	&	$+$63.0\phn	 & \phn 5.8 \\
5377.5490	&	0.3288	&	$+$68.8\phn	 &     20.0 \\
5387.6646	&	0.9810	&	$-$276.3\phn\phn & \phn 4.7 \\
5406.6203	&	0.0769	&	$-$216.8\phn\phn & \phn 4.6 \\
5409.6502	&	0.5717	&	$+$110.8\phn\phn & \phn 8.0 \\
5413.6381	&	0.2231	&	$+$87.2\phn	 & \phn 3.3 \\
5429.6113	&	0.8319	&	$-$166.1\phn\phn &     10.0 \\
5430.5901	&	0.9917	&	$-$262.7\phn\phn & \phn 7.8 \\
5444.5214	&	0.2670	&	$+$57.2\phn	 & \phn 3.4 \\
5446.5006	&	0.5903	&	$+$108.7\phn\phn & \phn 6.1 \\
\enddata
\end{deluxetable}

\begin{deluxetable}{lc} 
\tablewidth{0pc} 
\tablecaption{Orbital Elements\label{tab2}} 
\tablehead{ 
\colhead{Element}                      & \colhead{Value}       } 
\startdata 
$P$ (d)                       \dotfill & 6.1228\tablenotemark{a} \\ 
$T$ (HJD -- 2,400,000)        \dotfill & $54622.43 \pm  0.25$     \\ 
$e$                           \dotfill & $0.19 \pm 0.06$          \\
$\omega$ (deg)                \dotfill & $197 \pm 14$             \\
$K_1$ (km s$^{-1}$)           \dotfill & $230.0 \pm 12.7$         \\ 
$V_0$ (km s$^{-1}$)           \dotfill & $-23.9 \pm 8.4$          \\ 
$a_1\sin i$ ($R_\odot$)       \dotfill & $27.3 \pm 1.5$           \\ 
$f(M)$ ($M_\odot$)            \dotfill & $7.3 \pm 1.2$            \\ 
r.m.s. (km s$^{-1}$)          \dotfill & 29.3                     \\ 
\enddata 
\tablenotetext{a}{Fixed to value obtained from analysis of ASAS photometry (\S 2).} 

\end{deluxetable} 

\begin{deluxetable}{ccccc}
\tablecaption{\ion{He}{1} $\lambda$5876 Measurements
}
\tablewidth{0pt}
\centering
\tablehead{
  \colhead{HJD}         &
  \colhead{}  &
  \colhead{}  &  
  \colhead{$W_{\lambda}$\tablenotemark{a} } &
  \colhead{$V_r$} \\  

  \colhead{$-$2,400,000}         &
  \colhead{Phase}  &  
  \colhead{Spectrograph} &
  \colhead{(\AA)} &  
  \colhead{(km s$^{-1}$)}}  

\startdata
       54552.698  &        0.611  &    R-C    &        $-$9.06  &        $-$31.3\phn \\ 
       54587.861  &        0.354  &    R-C    &        $-$8.20  &        $-$36.0\phn \\ 
       54600.788  &        0.465  &    R-C    &        $-$9.70  &        $+$2.2 \\ 
       54623.846  &        0.231  &   echelle   &        $-$8.13  &        $-$13.8\phn \\
       54624.835  &        0.392  &   echelle   &        $-$6.23  &        $-$14.2\phn \\
       54626.844  &        0.721  &   echelle   &        $-$8.40  &        $-$6.7 \\
       54627.838  &        0.883  &   echelle   &        $-$8.39  &        $+$12.7\phn \\
       54638.821  &        0.677  &    R-C    &        $-$10.02\phn &        $-$4.5 \\ 
       54642.773  &        0.322  &    R-C    &        $-$8.91  &        $+$8.9 \\ 
       54647.788  &        0.141  &    R-C    &        $-$10.50\phn &        $-$22.5\phn \\ 
       54652.770  &        0.955  &    R-C    &        $-$9.27  &        $-$5.6 \\ 
       54682.716  &        0.846  &    R-C    &        $-$10.97\phn &        $+$4.0 \\ 
       54690.703  &        0.150  &    R-C    &        $-$10.40\phn &        $-$30.3\phn \\ 
       54902.828  &        0.795  &    R-C    &        $-$10.48\phn &        $-$6.1 \\ 
       54926.868  &        0.722  &    R-C    &        $-$11.42\phn &        $-$12.6\phn \\ 
       54984.862  &        0.194  &    R-C    &        $-$9.91  &        $-$38.8\phn \\ 
       54991.812  &        0.329  &    R-C    &        $-$8.28  &        $-$17.6\phn \\ 
       55020.802  &        0.063  &    R-C    &        $-$9.95  &        $-$2.3 \\ 
       55070.652  &        0.205  &    R-C    &        $-$10.08\phn &        $-$24.6\phn \\ 
       55111.530  &        0.882  &    R-C    &        $-$9.59  &        $+$10.6\phn \\ 
       55124.516  &        0.002  &    R-C    &        $-$9.05  &        $+$2.3 \\ 
       55325.864  &        0.887  &    R-C    &        $-$9.51  &        $+$9.2 \\ 
       55327.853  &        0.212  &   echelle   &        $-$9.76  &        $-$17.3\phn \\
       55335.854  &        0.519  &   echelle   &        $-$8.03  &        $-$1.2 \\
       55339.805  &        0.164  &   echelle   &        $-$9.60  &        $-$21.8\phn \\
       55346.866  &        0.317  &    R-C    &        $-$9.48  &        $-$23.9\phn \\ 
       55348.875  &        0.646  &   echelle   &        $-$8.77  &        $-$3.3 \\
       55358.505  &        0.218  &    R-C    &        $-$9.72  &        $-$17.6\phn \\ 
       55377.549  &        0.329  &   echelle   &        $-$7.89  &        $-$14.0\phn \\
       55378.634  &        0.506  &   echelle   &        $-$8.81  &        $-$5.4 \\
       55387.665  &        0.981  &   echelle   &        $-$9.37  &        $+$1.2 \\
       55385.828  &        0.681  &    R-C    &        $-$10.27\phn &        $-$12.3\phn \\ 
       55406.620  &        0.077  &   echelle   &        $-$8.01  &        $+$1.2 \\
       55429.611  &        0.832  &   echelle   &        $-$9.37  &        $+$1.8 \\
       55430.590  &        0.992  &   echelle   &        $-$9.36  &        $+$20.5\phn \\
       55444.521  &        0.267  &   echelle   &        $-$9.14  &        $-$12.3\phn \\
       55445.502  &        0.427  &   echelle   &        $-$6.92  &        $-$19.2\phn \\
       55446.501  &        0.590  &   echelle   &        $-$9.10  &        $-$5.6 \\
\enddata
\tablenotetext{a}{The $W_{\lambda}$ measurements include Na D emission and absorption, since the \ion{He}{1} and Na D features are blended in the lower resolution R-C spectra.}
\end{deluxetable}


\begin{thebibliography}{}
\bibitem[Borges Fernandes et al.(2001)]{borges}
	Borges Fernandes, M., de Ara\'{u}jo, F. X., Bastos Pereira, C., \& Codina Landaberry, S. J.
	2001, \apjs, 136, 747
\bibitem[Eggleton (1983)]{egg83}
        Eggleton, P. P.  1983, ApJ, 268, 368
\bibitem[Gies \& Bolton(1986)]{gies86}
        Gies, D. R., \& Bolton, C. T. 1986, \apj, 304, 371
\bibitem[Grundstrom et al.(2007)]{erika}
	Grundstrom, E. D., Gies, D. R., Hillwig, T. C., et al.
	2007, \apj, 667, 505
\bibitem[Hillier \& Miller(1998)]{cmfgen}
	Hillier, D. J., \& Miller, D. L.
	1998, \apj, 519, 354
\bibitem[Howell et al.(2006)]{how06}
	 Howell, S. B., Walter, F. M., Harrison, T. E., et al. 2006, \apj, 652, 709
\bibitem[Lanz \& Hubeny (2007)]{tlusty}
	Lanz, T., \& Hubeny, I.
	2007, \apjs, 169, 83
\bibitem[Lobel et al.(2012)]{lobel}
	Lobel, A., Groh, J., Torres, K., Gorlova, N., \& Martayan, C. 2012,
	in Four Decades of Research on Massive Stars, A Scientific Meeting in the Honour of Anthony Moffat, 
        ed. C. Robert, N. St-Louis, \& L. Drissen (San Francisco: ASP), in press
\bibitem[lopes (1992)]{lopes}
	Lopes, D. F., Damineli Neto, A., \& de Freitas Pacheco, J. A.
	1992, \aap, 261, 482
\bibitem[Manfroid et al. (1995)]{manfroid}
	Manfroid, J., Sterken, C., Cunow, B., et al.
	1995, \aaps, 109, 329
\bibitem[Marcolino et al.(2007)]{marcolino}
	Marcolino, W. L. F., de Ara\'{u}jo, F. X., Lorenz-Martins, S., \& Borges Fernandes, M.
	2007, \aj, 133, 489
\bibitem[Martins et al.(2005)Martins, Schaerer, \& Hillier]{mar05}
        Martins, F., Schaerer, D., \& Hillier, D.~J. 2005, \aap, 436, 1049	
\bibitem[Massey et al.(1981)]{mas81}
	Massey, P., Conti, P. S., \& Niemela, V. S.
	1981, \apj, 246, 145
\bibitem[Massey et al.(2007)]{mas07}
	Massey, P., McNeill, R. T., Olsen, K. A. G., et al. 2007, \aj, 134, 2474
\bibitem[Meynet (2003)]{mey03} 
	Meynet, G., \& Maeder, A. 
	2003, \aap, 404, 975
\bibitem[Morbey \& Brosterhus(1974)]{mor74} 
        Morbey, C. L., \& Brosterhus, E. B.\ 1974, \pasp, 86, 455
\bibitem[Muratorio et al.(2008)]{mur08}
	Muratorio, G., Rossi, C., \& Friedjung, M.
	2008, \aap, 487, 637
\bibitem[Nazarenko \& Glazunova(2006)]{na06}
        Nazarenko, V.~V., \& Glazunova, L.~V. 2006, Astr. Rep., 50, 369
\bibitem[North et al.(2007)]{north}
	North, J. R., Tuthill, P. G., Tango, W. J., \& Davis, J.
	2007, \mnras, 377, 415 
\bibitem[Pichardo et al.(2008)]{pich08}
	Pichardo, B., Sparke, L. S., \& Aguilar, L. A.
	2008, \mnras, 391, 815
\bibitem[Pojma\'{n}ski (2002)]{asas}
	Pojma\'{n}ski, G. 
	2002, Acta Astron., 52, 397
\bibitem[Pojma\'{n}ski \& Maciejewski(2004)]{asas2}
	Pojma\'{n}ski, G., \& Maciejewski, G.
	2004, Acta Astron., 54, 153
\bibitem[Porter \& Rivinius (2003)]{pr03}
	Porter, J. M., \& Rivinius, T. 
	2003, \pasp, 115, 1153
\bibitem[Scargle (1982)]{sca82}
        Scargle, J. D. 1982, \apj, 263, 835
\bibitem[Shafter et al.(1986)]{sha86}
        Shafter, A. W., Szkody, P., \& Thorstensen, J. R. 
	1986, \apj, 308, 765
\bibitem[Shore \& Brown(1990)]{sho90}
	Shore, S. N., \& Brown, D. N.
	1990, \apj, 365, 665
\bibitem[Shore et al.(1990)]{sea90}
	Shore, S. N., Brown, D. N., Bopp, B. W., et al. 1990, \apjs, 73, 461
\bibitem[Smith et al.(2011b)]{smi11b}
	Smith, N., Gehrz, R. D., Campbell, et al.
	2011b, arXiv:1105.2329
\bibitem[Smith et al.(2011a)]{smi11a}
        Smith, N., Li, W., Silverman, J. M., Ganeshalingam, M., \& Filippenko, A. V.
        2011a, \mnras, 415, 773
\bibitem[Smith \& Owocki (2006)]{so06} 
	Smith, N., \& Owocki, S. P. 
	2006, \apj, 645, L45
\bibitem[Sterken (1993)]{ste93}
	Sterken, C., Manfroid, J., Anton, K., et al. 
	1993, \aaps, 102, 79
\bibitem[Sterken (1995b)]{sterken95b}
	Sterken, C., Manfroid, J., Beele, D., et al.
	1995b, \aaps, 113, 31
\bibitem[Sterken (1995a)]{sterken95}
	Sterken, C., Stahl, O., Wolf, B., Szeifert, T., \& Jones, A.
	1995a, \aap, 303, 766
\bibitem[Tarasov (2000)]{tar00}
        Tarasov, A.~E. 2000, in ASP Conf. Ser. 214, The Be Phenomenon in Early-Type Stars, 
        ed. M.~A. Smith, H.~F. Henrichs, \& J. Fabregat (San Francisco: ASP), 644
\bibitem[Townsend \& Owocki (2005)]{to05}
	Townsend, R. H. D., \& Owocki, S. P. 
	2005, \mnras, 357, 251
\bibitem[Ud-Doula et al.(2009)]{udd09}
	Ud-Doula, A., Owocki, S. P., \& Townsend, R. H. D. 
	2009, \mnras, 392, 1022
\bibitem[van Genderen (2001)]{vg01}
	van Genderen, A. M.
	2001, \aap, 366, 508
\bibitem[Walborn \& Fitzpatrick(2000)]{wal00}
	Walborn, N. R., \& Fitzpatrick, E. L.
	2000, \pasp, 112, 50
\bibitem[Williams et al.(2009)]{wil09}
        Williams, S. J., Gies, D. R., Matson, R. A., \& Huang, W. 2009, \apj, 696, L137
\bibitem[Zhao et al.(2008)]{ming}
	Zhao, M., Gies, D., Monnier, J. D., et al.
	2008, \apj, 684, L95
\bibitem[Zucker (2003)]{zuc03} 
        Zucker, S.\ 2003, \mnras, 342, 1291
\end{thebibliography}
\end{document}